\documentclass[12pt]{scrartcl}
\usepackage{a4}
\usepackage{amsthm}
\usepackage{amsmath}
\usepackage{amssymb}
\usepackage{amsfonts}
\usepackage{mathrsfs}
\usepackage{dsfont}
\usepackage{latexsym}
\usepackage{color}
\usepackage{longtable}
\usepackage{bbm,exscale}
\definecolor{Myblue}{rgb}{0,0,0.6}
\usepackage[a4paper,colorlinks,citecolor=Myblue,linkcolor=Myblue,urlcolor=Myblue,pdfpagemode=None]{hyperref}
\usepackage[square,numbers,sort&compress]{natbib}
\usepackage[all,cmtip]{xy}

  \vfuzz \hfuzz
\newcommand{\E}{\text{e}}
\newcommand{\I}{\text{i}}
\newcommand{\C}{\mathds{C}}

\newcommand{\Z}{\mathds{Z}}
\newcommand{\MF}{\operatorname{MF}}
\newcommand{\Jac}{\operatorname{Jac}}

\def\1{\ifmmode\mathrm{1\!l}\else\mbox{\(\mathrm{1\!l}\)}\fi}
\newcommand{\one}{\mathbbm{1}}
\newcommand{\be}{\begin{equation}}
\newcommand{\ee}{\end{equation}}
\newcommand{\bes}{\begin{equation*}}
\newcommand{\ees}{\end{equation*}}

\newcommand{\qdef}{Q_{\rm def}}
\def\lra{\longrightarrow}
\def\n{n}%{\underline n} 
\def\m{m}%{\underline m} 
\def\rr{r}%{\underline r} 

\def\bB{\overline{B}{}}
\def\Sh{S^\#} \def\Bh{B^\#} \def\bBh{\overline{B}^\#}
\def\behk{\beta^{\#\,(k+1)}}  \def\bek{\beta^{(k+1)}}
\def\ukp{{}^{(k+1)} } \def\uk{{}^{(k)} } 

\def\lra{\longrightarrow}
\def\maxm{{\mathfrak m}}
\def\qd#1{Q_{(#1)}}
\def\ob#1{\text{\textsl{ob}}_{(#1)}}  \def\obo{\text{\textsl{ob}}}
\newcommand{\Ext}{\operatorname{Ext}}
\newcommand{\ext}{\operatorname{Ext}}
\newcommand{\Hom}{\operatorname{Hom}}
\def\id{{\rm id}} \def\im{{\rm im}}

\allowdisplaybreaks

\deffootnote[1em]{1em}{1em}{\textsuperscript{\thefootnotemark}}

\numberwithin{equation}{section}

\begin{document}

\title{Algorithmic deformation \\ of matrix factorisations}

\author{Nils Carqueville$^1$ \quad Laura Dowdy \quad Andreas Recknagel$^2$
\\
\\[-0.1cm] 
{\normalsize\slshape $^1$LMU M\"unchen Theresienstra\ss e~37, D-80333 M\"unchen, }\\[-0.1cm]
{\normalsize\slshape  \;\;\, Universe Cluster, Boltzmannstra\ss e~2, D-85748 Garching}\\[-0.05cm]
  {\normalsize $^2$\textsl{King's College London, Department of Mathematics,}} \\ [-0.1cm]
   {\normalsize \textsl{Strand, London WC2R\,2LS, UK}}\\[-0.1cm]
}

\date{}
\maketitle

\footnote{\tt \href{mailto:nils.carqueville@physik.uni-muenchen.de}{nils.carqueville@physik.uni-muenchen.de}, \tt \href{mailto:dowdy.laura@googlemail.com}{dowdy.laura@googlemail.com}, \tt \href{mailto:andreas.recknagel@kcl.ac.uk}{andreas.recknagel@kcl.ac.uk}}

\begin{abstract}
\noindent Branes and defects in topological Landau-Ginzburg models are described by matrix factorisations. We revisit the problem of deforming them and discuss various deformation methods as well as their relations. We have implemented these algorithms and apply them to several examples. Apart from explicit results in concrete cases, this leads to a novel way to generate new matrix factorisations via nilpotent substitutions, and to criteria whether boundary obstructions can be lifted by bulk deformations. 
\end{abstract}

\newpage

\tableofcontents

\section{Introduction}\label{sec:introduction}

Perturbative superstring theory at microscopic length scales involves the study of (deformations of) 
two-dimensional $\mathcal N=2$ superconformal field theories (CFTs) on the worldsheet of the 
strings. These CFTs describe string vacua, and amplitudes in string theory are computed as 
integrals over CFT correlators. 

The more exhaustive our knowledge of CFTs is, the better our understanding of string theory will be. 
Highly symmetric, so-called rational CFTs are under good control, both conceptually and computationally. 
But for theories with lower symmetry or in the case of, say, boundary conditions in rational CFTs that 
do not preserve the full extended symmetry algebra, the situation is much less satisfactory. Hence in 
order to get to grips with a wider class of string vacua, additional insight is needed. 

One source of such ideas is the correspondence between CFTs and Landau-Ginzburg models. The latter are 
non-conformal $\mathcal N=2$ supersymmetric field theories, and many interesting CFTs can be described as infrared fixed points of the renormalisation group flow of Landau-Ginzburg models. There is already 
overwhelming evidence for the CFT/LG correspondence at the level of full quantum field theories, 
and the picture becomes even clearer when one restricts to the topologically twisted sectors, which 
on the CFT side describes chiral primary fields and BPS branes. The practical value of the CFT/LG 
correspondence lies in the fact that several quantities of interest are protected under renormalisation 
group flow, thus they may be studied on the Landau-Ginzburg side, which is much less sensitive to 
loss of symmetry as may result from deformations. 

\medskip

A central topic in many parts of mathematical physics and in string theory in particular is that of 
moduli spaces. In the present paper we will discuss, interrelate and implement several algorithmic 
methods to study moduli spaces of boundary conditions in topologically twisted Landau-Ginzburg models, corresponding to deformations of BPS branes in $\mathcal N=2$ superconformal field theories. 

Twisted Landau-Ginzburg models are fairly well-understood examples of two-dimensional topological field 
theories~\cite{l0010269,ms0609042}, and all their structure is encoded in the 
superpotential~$W\in\C[x_1,\ldots,x_n] \equiv \C[x]$, where~$n$ is the number of chiral superfields.
In particular, the bulk sector (corresponding to bulk chiral primaries in the associated CFT) is given by the Jacobi ring $\operatorname{Jac}(W) = \C[x]/(\partial W)$~\cite{v1991}, and boundary conditions (corresponding to half-BPS branes) are described by matrix factorisations~$Q$ of~$W$~\cite{kl0210,bhls0305,l0312}. This means that $Q = (\begin{smallmatrix}0&q_1\\q_0&0\end{smallmatrix})$ is an odd supermatrix with polynomial entries such that $Q^2 = W\cdot \one$. Open strings starting on~$Q$ and ending on~$Q'$ are described by elements in the cohomology of the associated boundary BRST operator. It acts on homogeneous 
supermatrices~$\psi$ of the appropriate size as $\psi \longmapsto Q'\psi - (-1)^{|\psi|} \psi Q$, 
and we denote its even and odd cohomologies (describing ``bosonic'' and ``fermionic'' open strings) by $\operatorname{Ext}^2(Q,Q')$ and $\operatorname{Ext}^1(Q,Q')$, respectively. Matrix factorisations together with these cohomologies as morphisms form the $\Z_2$-graded D-brane category $\MF(W)$. 

\smallskip

It is easily explained what it means to deform a given boundary condition $Q=Q(x)$: we want to find matrix factorisations $Q_{\text{def}}(x;u)$ such that
\be\label{Qdefcond}
Q_{\text{def}}(x;0) = Q(x) \quad \text{and} \quad Q_{\text{def}}(x;u)^2 = W(x) \cdot \one
\ee
in the presence of the boundary moduli~$u$. (Later on, we will also allow simultaneous bulk deformations.) 
A perturbative treatment of this deformation problem immediately shows that first-order deformations are 
given by fermionic strings, i.\,e.~one adds representatives of elements in $\operatorname{Ext}^1(Q,Q)$ 
to~$Q$, but in general there will be (higher order) obstructions. These obstructions have an interpretation as the F-term equations of the effective low-energy field theory associated to the brane~$Q$. The F-terms are the physically interesting information that can be extracted from the effective superpotential $\mathcal W_{\text{eff}}$ as $\partial \mathcal W_{\text{eff}}=0$. By solving the deformation problem~\eqref{Qdefcond} we can obtain the F-terms without having to construct $\mathcal W_{\text{eff}}$ first. 

There are at least three different algorithmic methods to tackle the deformation problem~\eqref{Qdefcond}. 
For example, one may choose not to rely on any additional machinery and try for a direct construction of 
$Q_\text{def}$ and the obstructions, by solving the perturbative deformation equations order by order. 
For cases where obstructions do not arise this was already explained in~\cite{hw0404196}. With a little 
extra input from commutative algebra, one can also treat obstructions in this direct approach and 
obtain a computer-implementable algorithm. 

A more conceptual approach rests on the observation that deformations are also at the heart of topological string theory. Indeed, as explained e.\,g.~in~\cite{hll0402} string amplitudes are deformations of field theory correlators. On the other hand, any topological string theory can be completely described in terms of certain $A_\infty$-algebras~\cite{hll0402,c0412149}, see~\cite{c0904.0862,ck1104.5438} for the case of Landau-Ginzburg models. Thus it comes as no surprise that the deformation problem~\eqref{Qdefcond} can be phrased and solved in the language and with the methods of $A_\infty$-algebras. 

Finally, any deformation problem can be dealt with using the general machinery of deformation functors~\cite{Schless1968}. Luckily, at least in the case of matrix factorisations, this approach is also very explicit~\cite{Laudal1983,Siqveland2001}, and (given enough computing power) it is guaranteed to produce all solutions to~\eqref{Qdefcond}. 

\medskip

In this paper we will set up and review all these methods in detail, and we shall elucidate and expand on the connections between them. In a tiny nutshell, these connections are as follows: the condition $Q_\text{def}^2=W\cdot\one$ is the same as the Maurer-Cartan equation of the $A_\infty$-structure underlying topological string theory, and the Massey products featuring in the deformation functor approach are to be thought of as special kinds of $A_\infty$-products or amplitudes. 

We have also implemented from scratch (two versions of) the direct deformation algorithm mentioned above, as well as the algorithm building on the deformation functor approach. These implementations were carried out in the computer algebra system Singular and are available along with documentation and detailed examples at~\cite{cdrMFdeform}. This allows to study deformations of branes in arbitrary (orbifolds of) Landau-Ginzburg models. Since matrix factorisations are the sole input, our code may also be of interest for problems in singularity theory, where matrix factorisations play an important role as well. 

\medskip

The rest of the present paper is organised as follows. In section~\ref{sec:algorithms} we discuss in detail the three deformation methods mentioned above, explain their connections, and put them in a slightly wider context of general deformation theory. We also briefly indicate the important points for our implementation~\cite{cdrMFdeform} of the algorithms. These were applied to numerous examples, some of which (including a \textsl{family}~\eqref{MFfamily} of defects in a Landau-Ginzburg model corresponding to a rational CFT) we collect in section~\ref{sec:examples} for illustration. Playing with examples has also helped us to uncover some more general patterns. These are the subject of section~\ref{sec:observations}, where we discuss a construction of new matrix factorisations via ``nilpotent substitutions'', as well as criteria for when and how boundary obstructions can be lifted by bulk deformations. Some interesting open problems will be pointed out along the way.

\section{Deformation methods}\label{sec:algorithms}

In this section we spell out the details of the three approaches to our deformation problem mentioned in the introduction: trying to solve it with ``bare hands'', using $A_\infty$-algebras, and using the general theory of deformation functors and Massey products. Along the way it will also become clearer how these approaches are related, and we shall point out how we implemented two of the deformation algorithms in~\cite{cdrMFdeform}. 

\subsection{The direct method}\label{subsec:naive}

We begin by formulating the problem of constructing deformations of matrix factorisations in more detail. 
We will see how the 
fermionic and bosonic cohomologies control deformations and obstructions, respectively, 
and we will also show that a rather naive approach can be made into a fast 
algorithm which often allows to find families of deformations even if it may not yield 
the most general one. 

\smallskip
Given a matrix factorisation $Q$ of a Landau-Ginzburg potential $W$, i.\,e.\ 
$Q(x)^2 = W(x)\cdot\one_{2M}$ for some integer~$M$, one would like to find a family $\qdef(x;u)$ of matrix 
factorisations of $W(x)$ with $\qdef(x;0) = Q$ and such that the extra fermionic terms are 
polynomials or power series in the deformation parameters $u= (u_1,u_2,\ldots)$. We write 
$$
\qdef(x;u) = \sum_{n\geqslant 0}\;\qd{n}(x;u)
$$
where $Q_{(0)}=Q$ and where the term $\qd{n}$ consists of all order-$n$ monomials in the~$u_i$. One now tries to solve $\qdef^2 = W$  order by order. 

To first order in the $u_i$, one obtains the condition
$$
\qdef(x;u)^2 = Q^2 + \{Q,\qd{1}\} + {\cal O}(u^2) \stackrel{!}{=}  W\cdot \one
$$
which is satisfied iff $\qd{1}$ is an odd element in the kernel of the BRST operator, which we denote $d_Q$. 

The most straightforward, and most simple-minded ``solution'' to the deformation problem 
would now be to stick to $Q+\qd{1}$ with $\qd{1}\in\ker d_Q$, to treat the term $\qd{1}^2$ 
as an obstruction and to read off (quadratic) equations on the $u_i$ from each entry. 
This simplistic $\qdef$ would indeed provide a matrix factorisation of $W(x)$ as soon as 
the parameters are restricted to the zero locus of the obstruction equations. However, 
this typically leads to unnecessarily stringent conditions on the deformation parameters, 
which often only admit the trivial solution $u_i=0$. (Mapping cones are an exception.) 

One therefore aims at keeping the obstructions as ``small'' as possible by incorporating 
suitable higher order terms $\qd2,\qd3, \ldots$ into $\qdef$.  Consider the second order 
term of $\qdef^2 = W$, namely 
$$
\{Q,\qd2\} + \qd1^2  = 0\, .
$$
Once $\qd1$ has been chosen in the previous step, this is an equation for $\qd2$; it 
can be solved only if $\qd1^2 \in \im\, d_Q$ -- which is usually not the case. Let $\pi_B$ 
denote the projection of matrices to $\im\, d_Q$; then one can find $\qd2$ such that 
$$
\{Q,\qd2\} = - \pi_B\,\qd1^2
$$
and is left with a second order obstruction
$$
\ob2 := (\id-\pi_B) \qd1^2
$$
which cannot be removed by clever choices in $\qdef$. 

A simple computation shows that any component in $\qd1$ of the form $\{Q, \alpha\}$ 
can be absorbed into a redefinition of $\qd2$ -- thus the first order deformations are 
parametrised by the fermionic cohomology 
$H^1_Q=\ext^1(Q,Q)$, and 
we have $u=(u_1,\ldots,u_d)$ with $d=\dim H^1_Q$. This fits with the physical intuition 
that boundary deformations of branes arise from turning on boundary fields. 

Likewise, it is easy to see that the second order obstruction $\ob2$ is a bosonic element 
in $H^0_Q = \ext^2(Q,Q)$. 
This is in keeping with the standard pattern that deformations are 
controlled by $\ext^1$, obstructions by $\ext^2$. 

Proceeding to $n$-th order 
in the $u_i$, one arrives at the Maurer-Cartan type equation 
$$
\{Q,\qd{n}\} + \sum_{k=1}^{n-1} \qd{k}\qd{n-k} =  0\, ,
$$
and one ``solves'' this as before by choosing a $\qd{n}$ such that 
\be\label{SolveForQn}
\{Q,\qd{n}\} = - \pi_B\, \sum_{k=1}^{n-1} \qd{k}\qd{n-k} 
\ee
and adding the $n$-th 
order obstruction 
\be\label{ObstDirectAlg}
\ob{n} := (\id-\pi_B)\,\sum_{k=1}^{n-1} \qd{k}\qd{n-k} 
\ee
to the list collected so far. This obstruction need not be in $\ext^2(Q,Q)$ a priori 
(though it often happens to be in concrete examples, and this is always the case with 
the method of section~\ref{subsec:Massey}). One has the pretty relation 
$$
[\,Q,\ob{n}\,] = - \sum_{k=2}^{n-1}\;[\,\qd{n-k},\ob{k}\,]\, , 
$$
but for the left-hand side to vanish, one needs to choose the $\qd{k}$ 
such that the following additional condition is satisfied: 
\be\label{ObstinExt2}
\sum_{k=1}^{n-1}\,\{\qd{k},\qd{n-k}\} \in \ker d_Q \, .
\ee
At any rate, the algorithm allows to build up the deformed matrix factorisation 
and the total obstruction $\obo = \sum_{n\geqslant 2}\obo_{(n)}$ in 
\be
\qdef(x;u)^2 = W(x)\cdot \one + \obo(x;u) 
\ee
step by step. 
For many (but not all) cases of interest, the algorithm will terminate at finite 
order, and one is left with a polynomial $\qdef$ and polynomial obstructions 
\be
\obo(x;u) = \sum_j\, f_j(u)\,\phi_j\, .
\ee
If the obstructions are in $\ext^2(Q,Q)$, then the $\phi_j$ can be chosen as 
bosons representing a basis of $H^0_Q$; otherwise they are a basis of the complement 
of $\im\,d_Q$. 

Note that no matter how elementary or impenetrable the deformation algorithm, it 
is always straightforward to check correctness and whether one has reached a 
termination point, simply by squaring $Q+ \qd1+\ldots+ \qd{n}$ and comparing 
to $W + \ob2+\ldots+\ob{n}$. 

\medskip
Various comments are in order: 
The simple method described above is more or less the standard one given in~\cite{hw0404196}, 
except that obstructions terms are computed explicitly. 

Assuming that the algorithm terminates, one will obtain new matrix factorisations 
$\qdef(x;u)$ of $W(x)$, valid for 
\be
u \in L_f := \{\, u\in \C^d\;|\; f_j(u) = 0\,\} \, ,
\ee
the common zero locus of the $f_j(u)$ (which also has a physical interpretation 
as the flat directions of the effective superpotential ${\cal W}_{\rm eff}(u)$ on 
the topological brane described by $Q$ -- see section~\ref{subsec:Ainf} for further remarks). 
By analysing these new branes, one can make statements about renormalisation 
group flow patterns.

What is not guaranteed, however, is that the most general (``miniversal'') 
deformation is found: at each step, the obstruction terms are listed, but not 
solved, thus none of the abstract theorems  describing liftings and the associated 
obstruction theory (see section~\ref{subsec:Massey}) 
are applicable, because after the first obstruction has occurred, 
we are no longer dealing with a matrix factorisation of $W$. These issues are taken 
care of e.\,g.\ by Laudal's Massey product algorithm, which is accordingly far more 
intricate. 

The main advantage of the above direct method to compute deformations 
compared to the Massey product algorithm of section~\ref{subsec:Massey} is speed, 
making it possible to tackle a much wider class of topological branes than before. 
In examples which can be just about dealt with by the latter, the simpler 
algorithm from above is typically faster by a factor $10^5$. It should also 
be noted that in these cases, the zero loci obtained from the two algorithms 
happen to coincide. 

\smallskip

Note that at every step, the algorithm respects the grading by R-charges, therefore it 
applies to orbifolds of Landau-Ginzburg models without change. Moreover, one can 
incorporate bulk deformations of the potential $W(x)$, by solving 
\be\label{bulkQdef}
Q_{\text{def}}(x;u,t)^2 = \Big(W(x) + \sum_k t_k \,\varphi^{\text{B}}_k \Big)\cdot \one
\ee
where $\varphi^{\text{B}}_k\in \operatorname{Jac}(W)$ are bulk fields from the Jacobi ring 
of the Landau-Ginzburg potential. Mapping any such $\varphi^{\text{B}}_{k}$ to 
$\varphi^{\text{B}}_{k}\cdot\one$ produces a bosonic element of $\ker d_Q$. 
In case it is non-zero in the cohomology $H_Q^0$, one can find (certain) solutions to~\eqref{bulkQdef} by proceeding exactly as before if one imposes the equation $t_k = f_k(u)$ that relates the bulk and boundary parameters; in this case $\qdef(x;u,t)$ depends only implicitly on $t_{k}$. 
If, on the other hand, $\varphi^{\text{B}}_{k}\cdot\one = \{Q,\alpha_{k}\}$ for some 
fermionic matrix $\alpha_{k}$, then the only way to introduce a $t_{k}$-dependence 
into the deformed matrix factorisation is to add $t_{k}\,\alpha_{k}$ to 
the first order term $\qd{1}$. The algorithm can be carried out as before, except 
that (to our knowledge) it is no longer guaranteed that the higher order terms 
$\qd{n-1}$ can always be chosen such that~\eqref{ObstinExt2} is satisfied, i.\,e.\ 
such that the obstructions are always in $\ext^2(Q,Q)$. In the case of pure boundary deformations, the Massey product algorithm proves that such choices exist in principle. 

\smallskip

The procedure described above can be implemented on a computer in a straightforward way. 
The main steps at each order are first to compute the obstruction~\eqref{ObstDirectAlg} and then to 
solve the linear equation~\eqref{SolveForQn} for $Q_{(n)}$. Both steps boil down to a Gr\"obner basis 
computation; in the algebra package Singular one can basically use the functions \texttt{reduce} and 
\texttt{lift} to take these two steps, respectively. We have implemented this deformation algorithm 
as the function \texttt{deform} in our library \texttt{MFdeform.lib} available at~\cite{cdrMFdeform}. This comparably fast procedure is not guaranteed to make choices such that all obstructions are in $\Ext^2(Q,Q)$ (though often this is the case). One may also try to insist that such choices are made in the framework of the simple approach described above (as opposed to the setting of section~\ref{subsec:Massey} where we always have $\obo\in\Ext^2(Q,Q)$). This leads to a more involved construction implemented as the function \texttt{deformForceExt2} (see the documentation of~\cite{cdrMFdeform} for details). 
It manages to satisfy the conditions~\eqref{ObstinExt2} more often than \texttt{deform}, but if it does not one has to resort to another deformation method.

\subsection[$A_{\infty}$-products]{$\boldsymbol{A_{\infty}}$-products}\label{subsec:Ainf}

Another method to study deformations is using $A_{\infty}$-algebras. There are two related ways such algebras show up in our setting: on the one hand, given an appropriate $A_{\infty}$-structure one can immediately write down solutions to the deformation problem~\eqref{Qdefcond}, and on the other hand $A_{\infty}$-products are intimately related to Massey products which play a central role in the deformation algorithm discussed in section~\ref{subsec:Massey}. We shall now review these links after recalling the basic notions from $A_{\infty}$-theory. 

An \textsl{$A_{\infty}$-algebra} is a graded vector space $A=\bigoplus_{i}A_{i}$ with linear maps $r_{n}: A[1]^{\otimes n}\longrightarrow A[1]$ of degree $+1$ for all $n\in\Z_{+}$, where $A[1]$ is the same vector space~$A$ but with components $A[1]_{i}=A_{i+1}$. The \textsl{$A_{\infty}$-products}~$r_{n}$ have to satisfy the quadratic constraints
\be\label{A.inf.rel}
\sum_{k=1}^n r_{k}  r_{n}^k = 0
\ee
where we use the notation $r_{n}^k = \sum_{j=0}^{k-1} 1^{\otimes j}\otimes r_{n-k+1} \otimes 1^{\otimes(k-j+1)}$. $A_{\infty}$-products describe amplitudes in open topological string theory~\cite{hll0402, c0412149} where the constraints~\eqref{A.inf.rel} arise from Ward identities. 

Note that for $n=1$ and $n=2$ the $A_{\infty}$-relations~\eqref{A.inf.rel} tell us that $d=r_{1}$ is a differential: it squares to zero, $d^2=0$, and satisfies the Leibniz rule $d(a\cdot b)=d(a)\cdot b + (-1)^{|a|}a\cdot d(b)$ with respect to the product $a\cdot b=(-1)^{|a|+1}r_{2}(a\otimes b)$. The case $n=3$ in~\eqref{A.inf.rel} says that this product is associative up to a homotopy given by $r_{3}$. Thus we find that an $A_{\infty}$-algebra with $r_{n}=0$ for all $n\geqslant 3$ is precisely the same as a differential graded (DG) algebra. 

Given two $A_{\infty}$-algebras $(A,r_{n})$ and $(A',r'_{n})$ an \textsl{$A_{\infty}$-morphism}~$F$ is a family of linear maps $F_{n}:A[1]^{\otimes n}\longrightarrow A'[1]$ of degree~$0$ for all $n\in\Z_{+}$ compatible with the $A_{\infty}$-structures in the sense that
\be\label{A.inf.map}
\sum_{k=1}^n r'_{k}  F_{n}^k = \sum_{k=1}^n F_{k}  r_{n}^k
\ee
where $F_{n}^k = \sum_{j_{1}+\ldots+j_{k}=n} F_{j_{1}}\otimes\ldots\otimes F_{j_{k}}$. We call $F$ an \textsl{$A_{\infty}$-isomorphism} if $F_{1}$ is an isomorphism, and~$F$ is called an \textsl{$A_{\infty}$-quasi-isomorphism} if $F_{1}$ induces an isomorphism between the cohomologies $H_{r_{1}}(A)$ and $H_{r'_{1}}(A')$. 

The most fundamental result~\cite{k0504437, m9809} in $A_{\infty}$-theory is the \textsl{minimal model theorem}. The variant that we will use is as follows: for any DG algebra $(A,r_{n})$ its cohomology $H=H_{r_{1}}(A)$ can be endowed with $A_{\infty}$-products $\widetilde r_{n}$ (unique up to $A_{\infty}$-isomorphisms) such that there is an $A_{\infty}$-quasi-isomorphism $F:(H,\widetilde r_{n})\longrightarrow (A,r_{n})$. 

It will be useful to know some details of the construction. Let us decompose $A\cong H\oplus B \oplus L$ where $B=\operatorname{im}(r_{1})$ and~$L$ is the complement of $\operatorname{ker}r_{1}$. It follows that $r_{1}|_{L}:L\longrightarrow B$ is an isomorphism, and we can define the \textsl{propagator} $G=(r_{1}|_{L})^{-1} \pi_{B}$ where here and below we denote the projection to a subspace~$V$ by $\pi_{V}$. Note that $G^2 = 0$. Next we set $\lambda_{2}=r_{2}$ and then recursively
$$
\lambda_{n} = -r_{2} (G\lambda_{n-1}\otimes \id) - r_{2}  (\id\otimes G\lambda_{n-1}) - \sum_{\genfrac{}{}{0pt}{}{i,j\geqslant 2,}{i+j=n}} r_{2}  (G\lambda_{i} \otimes G\lambda_{j} ) \, . 
$$
Finally, the new $A_{\infty}$-products on~$H$ are given by $\widetilde r_{n}=\pi_{H}\lambda_{n}$, and the $A_{\infty}$-quasi-isomorphism~$F$ has components $F_{1}:H\lhook\joinrel\longrightarrow A$ and $F_{n}=G\lambda_{n}F_{1}^{\otimes n}$ for $n\geqslant 2$. We remark that different decompositions $A\cong H\oplus B \oplus L$ may lead to different (but $A_{\infty}$-isomorphic) $A_{\infty}$-structures $(H, \widetilde r_{n})$.

We should keep in mind that for our purposes DG algebras $(A,r_{n})$ are given by the \textsl{off-shell} open string space of cubic string field theory with its BRST differential~$r_{1}$ and operator product expansion~$r_{2}$. In this setting the above minimal model construction of the \textsl{on-shell} $A_{\infty}$-algebra $(H, \widetilde r_{n})$ encoding the amplitudes of open topological string theory arises by computing Feynman diagrams in the string field theory $(A,r_{n})$, see~\cite{l0107162}. 

\medskip

We now turn to the relation between $A_{\infty}$-algebras and deformation theory. (For a more complete study in the framework of topological string theory we refer to~\cite{ck1104.5438}.) Recall that we want to deform a matrix factorisation~$Q$ of~$W$, i.\,e.~we want to find all odd matrices $\delta_{Q}$ such that $(Q+\delta_{Q})^2=W\cdot \one$. Since~$Q$ already squares to~$W\cdot \one$ this condition is equivalent to
\be\label{MC1}
[ \,Q, \delta_{Q}\, ] + \delta_{Q}^2 = 0 \, .
\ee
If we write $\delta_{Q}=\sum_{n\geqslant 1}Q_{(n)}$ and introduce parameters $u_{i}$ by setting $Q_{(1)}=\sum_{i} u_{i} \psi_{i}$ where the $\psi_{i}$ are the ``directions'' into which we deform, then~\eqref{MC1} becomes
\be\label{MC}
[ \,Q, Q_{(n)} \,] + \sum_{j=1}^{n-1} Q_{(j)} Q_{(n-j)} = 0
\ee
for all $n\geqslant 2$. 

Now we remember that the off-shell open string space~$A$ of all matrices of the same size as~$Q$ together with the graded commutator $r_{1}=[Q, \,\cdot\,]$ as differential and matrix multiplication $r_{2}$ is a DG algebra. For any DG algebra, the equation $r_{1}(\delta) + r_{2}(\delta\otimes\delta) = 0$ with $|\delta|=1$ is called its Maurer-Cartan equation. Thus we find that our deformation problem~\eqref{MC} is precisely the problem of solving the Maurer-Cartan equation of the off-shell open string algebra. 

This is more than just introducing new vocabulary because we can relate the off-shell Maurer-Cartan equation to another physical condition coming from the on-shell open string space. To do this we first recall that for any $A_{\infty}$-algebra $(A,r_{n})$ its \textsl{Maurer-Cartan equation} takes the form $\sum_{n\geqslant 1} r_{n}(\delta^{\otimes n}) = 0$. An important result of~\cite{k9709040,m0001007} is that $A_{\infty}$-quasi-isomorphic $A_{\infty}$-algebras have isomorphic spaces of solutions to their Maurer-Cartan equations modulo gauge transformations. 

As a special case it follows that in this sense our deformation problem~\eqref{MC1} is equivalent to that of solving the Maurer-Cartan equation
\be\label{MC2}
\sum_{n\geqslant 2} \widetilde r_{n}(\widetilde \delta^{\otimes n}) = 0
\ee
of the on-shell $A_{\infty}$-algebra $(H, \widetilde r_{n})$. If the latter has a Calabi-Yau property which in particular means that the $A_{\infty}$-products $\widetilde r_{n}$ are cyclic with respect to the topological two-point correlator $\langle\,\cdot\,,\,\cdot\,\rangle$ of the boundary Landau-Ginzburg model (see~\cite{hll0402, c0904.0862} for a detailed discussion), then~\eqref{MC2} can be rewritten as the \textsl{F-term equation}
$$
\partial \mathcal W_{\text{eff}} = 0
$$
for the \textsl{effective D-brane superpotential}
$$
\mathcal W_{\text{eff}}(u) = \sum_{n\geqslant 2} \frac{1}{n+1} \,\Big\langle \psi_{i_{0}} , \widetilde r_{n}(\psi_{i_{1}}\otimes \ldots \otimes \psi_{i_{n}}) \Big\rangle \,  u_{i_{0}}  u_{i_{1}} \ldots  u_{i_{n}} \, . 
$$

We have thus found that solving $Q_{\text{def}}^2=W$ is the same as solving $\partial \mathcal W_{\text{eff}} = 0$. With the minimal model construction in hand one can also write down explicit solutions and obstruction constraints to~\eqref{MC}. For this we start by setting $Q_{(n)}=F_{n}(Q_{(1)}^{\otimes n})$ where $F: (H, \widetilde r_{n}) \longrightarrow (A,r_{n})$ is our minimal model $A_{\infty}$-quasi-isomorphism. To see whether this solves~\eqref{MC} we note that this equation can be written as
\be\label{MCr}
r_{1}(Q_{(1)}^{\otimes n}) = - r_{2}(F_{n}^2(Q_{(1)}^{\otimes n})) \, ,
\ee
where $r_{1}=[Q, \,\cdot\,]$ and $F_{n}^2(\psi^{\otimes n}) = \sum_{j=1}^{n-1} F_{j}(\psi^{\otimes j}) \otimes F_{n-j}(\psi^{\otimes (n-j)})$ as introduced after equation~\eqref{A.inf.map}. 

If the right-hand side of~\eqref{MCr} lies in the image~$B$ of $r_{1}$, 
\be\label{A.inf.obs}
\big((\id- \pi_{B})r_{2} F_{n}^2\big)  (Q_{(1)}^{\otimes n}) = 0 \, , 
\ee
then we can apply the propagator~$G$ and integrate~\eqref{MCr} to obtain the next order 
$$
Q_{(n)} = - (G r_{2} F_{n}^2) (Q_{(1)}^{\otimes n}) = F_{n} (Q_{(1)}^{\otimes n}) \, ,
$$
where the second step follows from equation~\eqref{A.inf.map}. Thus we arrive at the fact that computing the on-shell $A_{\infty}$-structure (encoded in the minimal model $A_{\infty}$-map~$F$) provides a solution to the deformation problem~\eqref{MC}: the deformed matrix factorisation $Q+\sum_{n\geqslant 1}Q_{(n)}$ is given in terms of $Q_{(n)}=F_{n}(Q_{(1)}^{\otimes n})$, and the obstructions are given by~\eqref{A.inf.obs}. 

We saw that the problem of computing deformations is solved as soon as we know the $A_{\infty}$-structure on the on-shell open string space. We also saw that computing the latter essentially boils down to explicitly computing the propagator~$G$, i.\,e.~the inverse of the BRST operator. In practise this can become rather involved, hence the more pedestrian deformation method of section~\ref{subsec:naive} can be very useful. 

\medskip

As reviewed in the next subsection, another general algorithmic method to obtain all deformations of a given matrix factorisation features Massey products. Roughly, whenever these products are defined they are given by $A_{\infty}$-products. To make the statement precise we will first recall what Massey products are in the context of DG algebras, following~\cite{lpwz0606144}. 

Let~$A$ be a DG algebra with differential~$d$. For a $d$-closed element~$a$ in~$A$ we denote by $[a]$ its cohomology class in $H=H_{d}(A)$. If~$a$ is homogeneous we set $\bar a = (-1)^{|a|+1}a$. Given two classes $[a_{1}], [a_{2}]\in H$, their \textsl{length-2 Massey product} $\langle [a_{1}], [a_{2}] \rangle$ is defined to be the set $\{[a_{1}a_{2}]\}$ with only one element given by the usual product. 

The \textsl{length-$n$ Massey products} $\langle [a_{1}], \ldots, [a_{n}] \rangle$ for $n\geqslant 3$ are also sets that are defined recursively, but only if the lower-order Massey products $\langle [a_{i}], \ldots, [a_{j}] \rangle$ exist for $i<j$ and $j-i<n-1$, and if they contain the zero element $[0]$. Under these assumptions one can show that there is a set $\{ a_{i,j} \}_{(i,j)\in I_{n}} \subset A$, indexed by $I_{n}=\{ (i,j) \,|\, 0<i<j<n,\, j-i+1\geqslant n \}$, whose elements satisfy $[a_{i-1,i}]=[a_{i}]$ and $d(a_{i,j}) = \sum_{k=i+1}^{j-1}\bar a_{i,k} a_{k,j}$ whenever $i<j-1$. Such a set $\{ a_{i,j} \}$ is called a \textsl{defining system} for $[a_{1}],\ldots, [a_{n}]$, and one sets
$$
\langle [a_{1}], \ldots, [a_{n}] \rangle = \left\{ \Big[ \sum_{i=1}^{n-1} \bar a_{0,i} a_{i,n} \Big] \;\Big|\; \{ a_{i,j} \} \text{ is a defining system for } [a_{1}],\ldots, [a_{n}] \right\} \, . 
$$

One relation between Massey products and $A_{\infty}$-products is as follows. Earlier we saw that for any DG algebra $(A,r_{n})$ one can construct $A_{\infty}$-structures $(H,\widetilde r_{n})$ on cohomology $H=H_{r_{1}}(A)$. Now the result of~\cite[Thm.~3.1~\&~Cor.~A.5]{lpwz0606144} is that if the Massey product exists for $[a_{1}], \ldots, [a_{n}]\in H$, then
$$
(-1)^{1+\sum_{0 < i < (n + 1)/2}|a_{n+1-2i}|} \, \widetilde r_{n}([a_{1}]\otimes\ldots\otimes [a_{n}]) \in \langle [a_{1}], \ldots, [a_{n}] \rangle
$$
for all $A_{\infty}$-structures on~$H$.

\subsection{The Massey product algorithm}\label{subsec:Massey}

Of the methods discussed so far, the second is relatively abstract, but conceptually interesting 
as $A_\infty$-algebras are the structure governing all of open topological string theory. The 
method described in section~\ref{subsec:naive} gives a very pedestrian yet fast algorithm to produce some deformations 
of a given matrix factorisation, but it is not all clear in which cases it detects all possible 
deformations. 

In contrast, the algorithm described in this section allows to construct the most general deformation. 
It is based on very general and far-reaching theorems on deformation functors by Schlessinger 
\cite{Schless1968} and a concrete method for computing liftings and obstructions developed by Laudal 
\cite{Laudal1983}. Massey products appear as a by-product in the latter construction and lend 
their name to the algorithm even though their definition is not required to understand the procedure. 

In this section, we first very briefly sketch Schlessinger's concept of deformation functors 
(narrowed down to our situation and with many details suppressed). This is done mainly so as 
to put the relatively technical Massey product algorithm into a wider context. After that, we 
present the concrete steps of this algorithm, including some details that are not spelt out 
explicitly in the literature. 

\smallskip
In \cite{Schless1968}, Schlessinger studies functors of Artin rings and gives  
necessary and sufficient conditions for such a functor to admit a ``hull''. In our 
present context, the relevant Artin rings are $S/\maxm^N$ or quotients thereof, with 
$S= \C[\![u_1,\ldots,u_d]\!]$ and $\maxm = (u_1,\ldots,u_d)$ the maximal ideal of the 
power series ring $S$. (Division by $\maxm^N$ was already used in section~\ref{subsec:naive}, when 
we considered deformations up to order $N-1$ in the parameters $u_i$, and this 
``finite-dimensionality'' is indeed the property of Artin rings which is relevant here.) 

The covariant functors considered in \cite{Schless1968} assign, to an Artin ring $A$, 
a set $F(A)$ -- which can e.\,g.\ be an isomorphism class of modules over $\C[x]\otimes A$. 
(Recall that a matrix factorisation $Q = (\begin{smallmatrix}0&q_1\\q_0&0\end{smallmatrix})$ determines 
a module $\operatorname{coker}q_1$ over $\C[x]/(W)$, from which we can recover $q_0$ and $q_1$ via a free resolution.) 
The set $F(\C)$ contains just one element $\xi_0$, corresponding to the 
undeformed structure (the matrix factorisation $Q$ in our case), and the first order 
deformations $F(S/\maxm^2)$ actually form a finite-dimensional vector space (called 
the tangent space of $F$), which in our case corresponds to 
the~$d$ boundary fermions in $H^1_Q = \Ext^1(Q,Q)$ that can be turned on.  
The elements $\xi \in F(A)$ are called (infinitesimal) deformations of $\xi_0$ 
over $A$ -- and correspond to our $Q_{\rm def}(u)$, truncated at certain orders. 

One natural question about such functors is when they are easily manageable in the following 
sense: ideally, one would like to a have a complete local algebra $S_\infty$ such that 
$F \cong h_{S_\infty}$ where $h_{S_\infty}(\,\cdot\,) = \Hom(S_\infty,\,\cdot\,)$; 
in this (rare) situation $F$ is called pro-representable. Suboptimally (but more common 
in cases of interest), one would like to find a so-called \textsl{hull} (or \textsl{miniversal 
deformation}) for $F$, which is a ring $S_\infty$ such that the tangent 
spaces $F(S/\maxm^2)$ and $h_{S_\infty}(S/\maxm^2)$ are isomorphic and such that 
for any surjection $A' \lra A$ of Artin rings, there is a surjective 
map $h_{S_\infty}(A') \lra h_{S_\infty}(A) \times_{F(A)} F(A')$; then $S_\infty$ still contains 
complete information about the deformation functor $F$. (We refer to \cite{Schless1968} 
or appendix C of~\cite{GreuelLossen} for the definition of this canonical map.)

More precisely, a miniversal deformation consists of the local algebra $S_\infty$ 
(which may not be an Artin ring) together with 
an element $\xi_\infty$ in $\hat F(S_\infty)$, where $\hat F$ denotes the functor 
extended to complete local algebras in a straightforward way by taking projective 
limits. In our case, the hull is given by $S_\infty =S/I_\infty$ where $I_\infty$ denotes 
the ideal spanned by the obstructions (F-terms) of the deformation, and $\xi_\infty$ 
corresponds to the deformed matrix factorisation $Q_{\rm def}(u)$ with all orders included. 

Schlessinger's main theorem gives conditions which guarantee the existence of a hull, 
see \cite{Schless1968}, and the proof contains a ``semi-constructive'' procedure how to 
ascend along a chain of Artin rings $S_{k} \subset S_{k+1}$, where each $S_n$ is determined 
by an ideal $I_k$ as $S_k = S/I_k$; we will use $p_{k+1} : S_{k+1} \lra S_k$ 
to denote the natural projection. One needs to lift an 
$F(S_k)$-element $\xi_k$ (say, a deformed matrix factorisation $Q_{\rm def}(x;u)$ 
that squares to $W(x)$ up to terms of order $k+1$ in the $u_i$) 
to an element  $\xi_{k+1} \in F(S_{k+1})$. The main step in the process is: given 
$S_k = S/I_k$ and $\xi_k$, find the smallest ideal $I_{k+1}\subset S$ such 
that $\maxm\,I_k \subset I_{k+1} \subset I_k$ and such that there is a pre-image 
$\xi_{k+1}$ of $\xi_k$ under $F(p_{k+1})$. 

The first condition on $I_{k+1}$ means that $p_{k+1}$ is a \textsl{small extension}, 
i.\,e.\ a surjective  ring homomorphism such that $\maxm_{S_{k+1}}\cdot \ker p_{k+1} = 0$, 
where $\maxm_{S_{k+1}}$ denotes the maximal ideal of $S_{k+1}$. Small extensions are relatively 
easy to handle in practise (the canonical surjections $S/\maxm^{N+1} \lra S/\maxm^N$ are 
special examples, others will appear in the Massey product algorithm at each lifting step),
but they suffice to determine the behaviour of $F$ under arbitrary extensions, see 
\cite{Schless1968}. The second condition on $I_{k+1}$ has to be 
imposed because in the lifting of $\xi_k$ to the next step one may encounter obstructions, 
which need to be absorbed into $I_{k+1}$. 

\medskip 

Laudal's algorithm  \cite{Laudal1983} allows to construct the deformation functor 
as well as the obstruction ideals $I_k$ (and thus the Artin rings $S_k = S/I_k$) 
explicitly, and it yields the hull $S_\infty$ in the limit $k\to\infty$. 
We provide some details of the algorithm, adapted to deformations of matrix 
factorisations (Laudal's theorems have a somewhat wider scope). Apart from 
Laudal's work, the papers \cite{Siqveland2001} and \cite{KnappOmer} may 
be useful. 

To obtain explicit formulas, one works in a monomial basis 
$u^{\n} := u_1^{n_1}\cdots u_d^{n_d}$ 
with $\n\in (\Z_+)^d$. We can use this to expand the obstruction 
$\obo \in \ext^2(Q,Q) \otimes S_\infty$ in two ways, 
$$
\obo = \sum_j\, f^{(\infty)}_j(u)\,\phi_j = \sum_{\n}\, y(\n)\,u^{\n}
$$ 
where the boundary bosons $\phi_j$ are representatives of a basis of $H^0_Q=\ext^2(Q,Q)$, 
and where the bosonic matrices $y(\n)$ will be computed by Laudal's algorithm. 
Along the way, one also computes coefficients $\beta_{\n,\rr}$ such that there is 
a fermionic matrix $\alpha_{\rr}$ with 
$$
\{Q,\alpha_{\rr}\} = \sum_{\n}\,\beta_{\n,\rr}\,y( \n)\, .
$$
These pre-images then contribute the term $u^{\rr}\,\alpha_{\rr}$ to $Q_{\rm def}(u)$. 
In the language of the preceding section, the linear combination on the right-hand side is the 
projection $\pi_B$ of $y(\n)$ onto the image of $r_1$. 

The process starts out as in section~\ref{subsec:naive}: lift $\qd0 =Q$ over $\C$ to $\qd0+\qd1$ over 
$S_1 = S/\maxm^2$, which is unobstructed. At second order, obstructions $\ob2$ may appear, 
which can be brought into the form $\ob2 \in \ext^2(Q,Q) \otimes I_2$ and thus define an ideal 
$I_2\subset S$ and also the next Artin ring $S_2 = S/(\maxm^3 + I_2)$ in our chain of 
liftings. One can expand $\ob2$ into monomials to obtain $y(\n)$ for 
$|\n| := n_1 + \ldots + n_d \leqslant 2$, or into the $\ext^2$-basis $\phi_j$ 
with coefficients $f^{(2)}(u)$. 
A second order term $\qd2$ of the deformed matrix factorisation is 
found as a (non-unique) pre-image $\{Q,\qd2\} = (Q + \qd1)^2 - \ob2 + \mathcal O(u^3)$, 
and this pre-image fixes coefficients $\beta_{\n,\rr}$ for length-2 multi-indices.

Suppose $y(\n)$ and $\beta_{\n,\rr}$ have been constructed up to order $k$, which 
means that the rings $S_l$, as well as $\qd{l}$ and $\ob{l}$ are known for $1\leqslant l \leqslant k$. 
The ring $S_k$ has a finite basis (over $\C$) consisting of monomials 
$u^{\n}\,,\;\n\in \bB_k$ for some set $\bB_k  \subset (\Z_+)^d$; these 
monomials satisfy $|\n| \leqslant k$. 
The ideal $I_k$ is generated by polynomials $f\uk_j$, which consist of the 
first $k$ orders of the full F-terms, with $j=1,\ldots,{\rm dim}\ext^2(Q,Q)$. 

In order to find the obstruction ideal and deformed matrix factorisation at the next step $k+1$, 
consider the equation 
\begin{align}\label{qdefsq}
Q_{\rm def}^2 - W\cdot \one  
&= \sum_j f^{(k+1)}_j\;\phi_j + {\cal O}(u^{k+2})
\nonumber \\ 
&=  \sum_{|\m|\leqslant k+1}\ \sum_{{ \m_1,\m_2\in\bB_k, \atop  \vphantom{S^S}\m_1+\m_2=\m}} 
\alpha_{\m_1}\alpha_{\m_2}\; u^{\m} + \bigl\{ Q,\tilde\alpha \bigr\} + {\cal O}(u^{k+2}) \, .
\end{align}
The expression in the first line contains obstructions, the one in the second line 
is obtained from squaring the first $k$ terms of the deformed matrix factorisation. 
The precise form of the $\tilde\alpha$-term is of no interest for the following, 
as we will pass to the $Q$-cohomology directly; the resulting equation 
holds in ${\rm Ext}^2(Q,Q) \otimes S$. 

It still involves higher order terms ${\cal O}(u^{k+2})$, but we can project down to any 
$S_l$ or any other Artin ring: a useful choice is $\Sh_{k+1} = S/J_{k+1}$ with 
$J_{k+1} := \maxm^{k+2} + \maxm\,I_k$; we denote the projection by 
$p^\#_{k+1} : S^\#_{k+1} \lra S_{k+1}$. 

Computing in this auxiliary ring includes the next order in the $u_i$ while 
still taking the F-term equations from the previous order into account. 
A monomial basis $u^{\n}$ for $\Sh_l$ can then be labelled by 
$\n\in \bB^\#_l :=  \Bh_l \cup \bB_{l-1}$, where the $\n \in B^\#_l$ all have 
length $|\n| = l$. Using this basis and the generators $f_j^{(k)}$ of the ideal 
$I_k$, any monomial $u^{\m}$ with $|\m| \leqslant k+1$ has an expansion 
$$
u^{\m} = \sum_{\n \in \bBh_{k+1}}\behk_{\m,\n}\,u^{\n} + \sum_j\, \behk_{\m,j}\,f^{(k)}_j
$$
in the ring $\Sh_{k+1}$ with unique complex coefficients $\behk$.

This is already sufficient to determine the next order of the obstruction polynomials. 
To see this, project equation~\eqref{qdefsq} to cohomology classes $[\,\cdot\,]$, then apply 
${\rm id}_{\Ext^2} \otimes p^\#_{k+1}$. The higher order terms disappear, and 
we can use the basis of $\Sh_{k+1}$ to express the $u^{\m}$ using the $\behk_{n,m}$. 
This gives
$$
\bigl[\;\sum_j f^{(k+1)}_j\;\phi_j\bigr] = 
\Bigl[\ \sum_{\n\in \Bh_{k+1}}\  \sum_{{ \m_1,\m_2\in\bB_k, \atop  \vphantom{S^S}\m_1+\m_2=\m}}\ 
\behk_{\m,\n}\ \alpha_{\m_1}\alpha_{\m_2}\;u^{\n} \ \Bigr]
+ {\cal O}(u^{k+2})\, .
$$
We have also split up the summation over  $\n \in \bBh_{k+1} = \bBh_{k+1} \cup \bB_k$ into 
the new terms at order $k+1$ and lower order terms with $\n \in \bB_k$; the latter are all hidden 
in the ``remainder'' and contribute precisely the obstruction $\ob{k} = \sum_j\;f_j^{(k)}\phi_j$ 
already computed in the previous steps of the algorithm. 
One can now read off the new 
$$
y(\n) := \sum_{{ \m_1,\m_2\in\bB_k, \atop  \vphantom{S^S}\m_1+\m_2=\m}}
\behk_{\m,\n}\ \alpha_{\m_1}\alpha_{\m_2}
$$
and one arrives at the following expression for a representative of the obstruction up 
to order $k+1$:
\be\label{obstrep}
\sum_j\ f^{(k+1)}_j\;\phi_j = 
\sum_{\n \in \Bh_2  \cup \ldots \cup \Bh_k\cup \Bh_{k+1}} y(\n)\;u^{\n} \, .
\ee
In particular, the $f^{(k+1)}_j$ 
determine the new ideal $I_{k+1}$ and the next Artin ring $S_{k+1} = S/I_{k+1}$ in our 
sequence of liftings. 

It remains to find the terms of order $k+1$ for the deformed matrix factorisation $Q_{\rm def}(u)$. 
First note that $S_{k+1}$ has a monomial basis $u^{\n}$ with $\n\in\bB_{k+1} = \bB_k \cup B_{k+1}$ 
where $B_{k+1} \subset \Bh_{k+1}$, and one has a unique expansion 
$$
u^{\n} = \sum_{\m \in \bB_{k+1}} \bek_{\n,\m}\,u^{\m} 
$$
for any monomial $u^{\n}$ with $|\n| \leqslant k+1$ in $S_{k+1}$. 
(One can show that $\beta\ukp_{\m,\n} = \beta\uk_{\m,\n}$ as long as 
$|\m|, |\n| \leqslant k$, so the coefficients computed at lower 
orders carry over.) 

Next, we project \eqref{obstrep} to the $Q$-cohomology and apply 
${\rm id}_{\Ext^2} \otimes p_{k+1}$. This annihilates the left-hand side, and 
using the $\bek$ we obtain, for each  $\rr \in \bB_{k+1}$, a relation 
$$
 \Bigl[\ \sum_{\n \in B'_2 \cup B'_3 \cup \ldots \cup B'_{k+1}}
   \bek_{\n, r}\ y(\n)\ \Bigr] = 0 \, .
$$
Since the expression in brackets is zero in cohomology, it can be 
written as $\{Q,\alpha_{\rr}\}$ for some fermionic matrix $\alpha_{\rr}$. 
These form a defining system for the Massey products at order $k+1$ 
(see \cite{Laudal1983}) and, more importantly for us, make up the 
deformed matrix factorisation $Q_{\rm def}(u)$ up to this order. 

\medskip
The main difference between this algorithm and the pedestrian computation 
from section~\ref{subsec:naive} is that at each step, the obstructions are carefully 
quotiented out by restricting to the rings $S_k$ -- so the deformed 
matrix factorisation actually exists over this ring. Lifting this to 
small extensions $S_{k+1} \lra S_k$ (or $\Sh_{k+1} \lra S_k$) is then possible 
if and only if the obstruction $\ob{k+1}$ vanishes. (This is more or 
less obvious, but see \cite{Laudal1983}, \cite{Siqveland2001} 
or \cite{GreuelLossen} for details of the proof.) Again, any non-trivial 
obstruction arising in this step is then removed by dividing by $I_{k+1}$. 

Therefore, the Massey product algorithm guarantees that no obstruction 
is overlooked, and the end result is a deformation functor as introduced 
by Schlessinger, yielding $Q_{\rm def}(u)$,  together with a hull that 
describes all possible deformations of the original matrix factorisation 
and captures the F-terms of the associated topological brane. 

\smallskip

Our implementation of the above algorithm relies on the same basic Singular functions used also for the procedure described in section~\ref{subsec:naive}. However, the various auxiliary rings and their relations make the Massey product algorithm much more intricate. This subtle bookkeeping was carried out in programming the function \texttt{MFcohom\textunderscore def} in our library \texttt{MFMassey.lib} available at~\cite{cdrMFdeform}.

\section{Some examples}\label{sec:examples}

We have applied the algorithms of the preceding section to numerous examples and 
will discuss a few in the following. We will focus on deformed matrix factorisations 
and obstructions (F-terms), not on effective superpotentials. Not only is it a tedious 
task to find $\mathcal{W}_{\rm eff}$, but the result depends very much on the moduli 
space coordinates one chooses. (Note that any transformation $u_i \longmapsto 
\tilde u_i = M_{ij}u_j + p_i(u)$ with an invertible complex matrix $M_{ij}$ is allowed, 
with arbitrary polynomials $p_i \in \maxm^2$. Imposing R-charge conservation is 
in general insufficient to prevent wild mixings in higher order terms of 
$\mathcal{W}_{\rm eff}$.) 

Information that is invariant under transformations of the $u_i$ is given by the 
RG flow pattern: what are (up to equivalence) the matrix factorisations $\qdef(x;u)$ 
with $u\in L_f$ that the initial $Q$ can be deformed into? For the examples discussed 
below, we give that information where known, but so far we have to resort to case by 
case checks for equivalence (apart from an inequivalence criterion using Fitting 
ideals, see section~\ref{subsec:tensorproducts}). It would be very useful to find tools that are both general 
and easy to implement to tackle questions of (in-)equivalence of $\qdef(x;u)$ for 
different $u$. 

We start with minimal model examples; unsurprisingly, the zero loci of the obstructions 
and the RG flow patterns are not very rich, still we uncover some flows that were not 
given in the literature. Also, the examples show that the simple method from section~\ref{subsec:naive}
allows to treat cases in practise that were way out of reach for present implementations 
of the Massey product algorithm. 

In section~\ref{subsec:tensorproducts}, we consider branes for tensor products of minimal models. Here richer flow patterns emerge, and in particular we find a one-parameter family of inequivalent 
matrix factorisations in a Landau-Ginzburg model corresponding to a CFT which is 
rational with respect to its maximal symmetry algebra. These are to be interpreted 
as symmetry-breaking branes -- objects that are notoriously difficult to construct 
in CFT. 
We also add some very brief remarks on deformations of topological branes in $c=9$ 
Gepner models. Here, one often encounters the problem that deformation algorithms 
need not terminate, and one would need additional structural insights to uncover 
the full moduli space. Still, the method from section~\ref{subsec:naive} at least allows to 
compute F-terms to rather high order. 

Readers are invited to try out the programs made available at \cite{cdrMFdeform} at home.

\subsection{Minimal models}\label{subsec:minimal}

Minimal $\mathcal N=2$ CFTs have an ADE classification, and so do the corresponding Landau-Ginzburg models. There is not much to say about deformations of matrix factorisations of A-type potentials $W_{\text{A}_n}=x^{n+1}$. For small values of~$n$, the algorithms of section~\ref{subsec:naive} and~\ref{subsec:Massey} both produce only trivial deformations when applied to indecomposable objects in $\MF(W_{\text{A}_n})$: $u_i=0$ is the only solution to the obstruction equations. The method of section~\ref{subsec:naive} can also easily treat $n\gtrsim 100$, with the same result. 

The situation is similarly uninteresting for potentials $W_{\text{A}'_n}=x^{n+1}+z^2$ corresponding to A-type CFTs with the opposite GSO projection. Here the obstruction equations do admit non-trivial solutions, but the associated deformed matrix factorisations themselves are trivial. 

\medskip

Minimal models with D-type modular invariant correspond to Landau-Ginzburg models with 
potentials 
$W_{\textrm{D}_{n+2}} = x_1^{n+1} - x_1 x_2^2$. Their matrix factorisations 
have e.\,g.~been studied in \cite{kst0511155} and \cite{Brunner:2005pq}. We will use 
the notations of the latter paper, where the indecomposable matrix factorisations 
were labelled as follows: 
For $n$ odd, there is a single rank $M=1$ matrix factorisation 
$\mathcal R_0$, and for even $n$ two additional ones $\mathcal R_\pm$ exist; none of these allow for fermions, 
so they have no deformations (unless one considers direct sums). There are two sequences 
of rank 2 factorisations, called $\mathcal S_l$ and $\mathcal T_l$ in \cite{Brunner:2005pq}. 

In contrast to A-type minimal model, deformations often admit non-trivial zero loci for 
the obstructions, along with non-trivial RG flow patterns. We will sketch the results 
for two simple examples; in both of these, $\qdef(x;u)$ can be computed with our direct 
algorithm as well as the Massey product algorithm, and the results coincide. Using the direct method of section~\ref{subsec:naive}, it takes no more than minutes to compute deformations of indecomposable objects in $\MF(W_{\text{D}_n})$ for $n\lesssim 30$. 

For $n=3$, consider the matrix factorisation $Q=\mathcal S_1$ in the notation of \cite{Brunner:2005pq}. 
Deforming by the two fermions leads to  
$\qdef(x;u) = (\begin{smallmatrix}0&q_1\\q_0&0\end{smallmatrix})$ with
\begin{align*}
q_1 & =  \begin{pmatrix}
x_1+u_{1} & x_1x_2-u_{1}u_{2} \\
-x_2+u_{2} & -x_1^3+x_1^{2}u_{1}-x_1u_{1}^{2}+x_2u_{2}+u_{1}^{3} \\
\end{pmatrix} , \\
q_0 & =  \begin{pmatrix}
x_1^3-x_1^{2}u_{1}+x_1u_{1}^{2}-x_2u_{2}-u_{1}^{3} & x_1x_2-u_{1}u_{2} \\
-x_2+u_{2} & -x_1-u_{1}\\
\end{pmatrix}\vphantom{\sum^{N^N}} ,
\end{align*}
and the obstruction can be read off from 
$\qdef(x;u)^2 = (W_{\textrm{D}_5}(x) - (u_{1}^{4} +u_{1}u_{2}^{2}))\cdot \one$. 
The zero locus is 
$$ 
L_f = \{ (0,u_2)\} \cup \{(u_1,\pm u_1^{3\over2})\} \subset \C^2
$$
which can be reproduced as the flat directions of the effective superpotential 
${\cal W}_{\textrm{eff}}(u) = (u_1^4u_1+{1\over3}u_1u_2^3)^2 -{104\over99}u_1^{11}$. 
One can now evaluate $\qdef(x;u)$ at selected points in $L_f$ and check whether the 
resulting matrix factorisation of $W_{\textrm{D}_5}(x)$ is equivalent to one from the 
list of fundamental matrix factorisations. In the present example, we found that $\qdef(x;u)$ is trivial for all points we 
looked at: the brane $\mathcal S_1$ is not stable, but decays completely under any non-trivial 
perturbation. 

The case $n=4$ with $Q=\mathcal T_2$ displays a richer pattern of RG flows. This brane has 
four fermions and six bosons, and the Massey product algorithm yields $\qdef(x;u)$ 
built from the blocks
\begin{align*}
q_1 & =  \begin{pmatrix}
x_1^2+x_1u_{2}+u_{4} & x_2+x_1u_{1}+u_{3} \\
-x_2+x_1u_{1}+u_{3} & -x_1^2+x_1u_{2}+u_{1}^{2}-u_{2}^{2}+u_{4}
\end{pmatrix} , \\
q_0 & =  \begin{pmatrix}
x_1^{3} -x_1^{2}u_{2}-x_1u_{1}^{2}+x_1u_{2}^{2}-x_1u_{4}  & x_1x_2 +x_1^{2}u_{1}+x_1u_{3}  \\
-x_1x_2 + x_1^{2}u_{1}+x_1u_{3} & -x_1^{3} -x_1^{2}u_{2}-x_1u_{4}\\
\end{pmatrix} \vphantom{\sum^{N^N}} .
\end{align*}
The obstruction polynomials are 
\begin{align*}
f_1(u) & = u_{1}^2 u_{2}-u_{2}^3-2 u_{1} u_{3}+2 u_{2} u_{4}\, ,\\ 
f_2(u) & = -{1\over2} u_{1}^2 u_{2}^2+4 u_{1} u_{2} u_{3}-u_{1}^2 u_{4}+u_{2}^2 u_{4}+u_{3}^2-u_{4}
\end{align*}
and lead to a rather complicated zero locus $L_f$ with several intersecting branches. 
Selecting special points from $L_f$ and testing for equivalence with known D-type matrix 
factorisations, one finds that $\mathcal T_2$ can under perturbations flow to $\mathcal T_1$ with four 
bosons and two fermions (this happens e.\,g.\ for $u_1=u_2=1, u_3=u_4=0$), or to $\mathcal T_0$ 
with two bosons and no fermions (e.\,g.\ for $u_1=u_2=0, u_3=u_4=1$). These results 
complement the general statements on mapping cones established in~\cite{Brunner:2005pq}. 

\medskip

\begin{table}[t]
\begin{center}
\begin{tabular}{lllll} 
& matrix factorisation & size & terminates at order & time taken \\
\hline
\hline
$E_6$: & $Q_1$ & $8 \times 8$ & 15 & seconds \\
& $Q_2$ & $12 \times 12$ & 15 & seconds \\
& $Q_3$ & $8 \times 8$ & 12 & seconds \\
& $Q_4$ & $8 \times 8$ & 12 & seconds \\
& $Q_5$ & $4 \times 4$ & 12 & seconds \\
& $Q_6$ & $4 \times 4$ & 12 & seconds \\
\hline
$E_7$: & $Q_1$ & $8 \times 8$ & 23 & seconds \\
& $Q_2$ & $12 \times 12$ & 25 & seconds \\
& $Q_3$ & $16 \times 16$ & 24 & minutes \\
& $Q_4$ & $8 \times 8$ & 23 & seconds \\
& $Q_5$ & $12 \times 12$ & 23 & seconds \\
& $Q_6$ & $ 8\times 8$ & 22 & seconds \\
& $Q_7$ & $4 \times 4$ & 18 & seconds \\
\hline
$E_8$: & $Q_1$ & $8 \times 8$ & 44 & seconds \\
& $Q_2$ & $12 \times 12$ & 43 & seconds \\
& $Q_3$ & $16 \times 16$ & 40 & 1 hour \\
& $Q_4$ & $20 \times 20$ & 41 & 3 days \\
& $Q_5$ & $24 \times 24$ & ?? & ?? \\
& $Q_6$ & $ 12\times 12$ & 35 & minutes \\
& $Q_7$ & $16 \times 16$ & 37 & 4 hours \\
& $Q_8$ & $8 \times 8$ & 38 & seconds \\
\hline
\hline
\end{tabular}%
\caption{Deformations of indecomposable matrix factorisations for minimal models of type E, using the same numbering as in~\cite{kst0511155}. }
\label{ADE-E-table}
\end{center}%
\end{table}%

Deformed matrix factorisations for the three exceptional singularities
$W_{\textrm{E}_6} =  x_1^3 + x_2^4 - x_3^2$, $W_{\textrm{E}_7} =  x_1^3 + x_1x_2^3 - x_3^2$, 
$W_{\textrm{E}_8} =  x_1^3 + x_2^5 - x_3^2$ are much harder to construct. The Massey 
product algorithm can so far only tackle the two simplest E$_6$ matrix factorisations 
and the simplest one for E$_7$. The pedestrian algorithm from section~\ref{subsec:naive}, on the other 
hand, produces results (summarised in table~\ref{ADE-E-table}) for all the indecomposable matrix factorisations listed 
in~\cite{kst0511155} except for the biggest E$_8$ matrix factorisation~$Q_5$, where the algorithm does not terminate on the ordinary desktop computers we have access to.

\subsection{Tensor products of minimal models}\label{subsec:tensorproducts}

\subsubsection*{A family of defects}

We now turn to matrix factorisations of $x^d-y^d$. These may either be viewed as branes in the tensor product theory of two A-type models, or as defects between two such theories. 

Let us start with the simplest non-trivial case where $d=3$. The ADE classification of simply-laced complex semisimple Lie algebras and isolated singularities together with the isomorphism $\mathfrak{su}(2) \oplus \mathfrak{su}(2) \cong \mathfrak{so}(4)$ suggests that $\MF(x^3-y^3)$ should be equivalent to $\MF(x^2y-y^3)$. The latter category describes branes in B-twisted Landau-Ginzburg models of type D$_{4}$, and all its objects are known to be isomorphic to direct sums of a finite set of indecomposable objects, see e.\,g.~\cite{Yoshinobook}. One says that $\MF(x^2y-y^3)$ is of \textsl{finite type}. 

$\MF(x^3-y^3)$ is also of finite type, and we can obtain its indecomposables from those of $\MF(x^2y-y^3)$ via the equivalence $f_{*}:\MF(x^2y-y^3) \longrightarrow \MF(x^3-y^3)$. Note that this means that defects in the A$_{2}$-model are the ``same'' as branes in the D$_{4}$-model. The functor $f_{*}$ is induced by the isomorphism $f:\C[x,y]\longrightarrow \C[x,y]$ given by
$$
x \longmapsto \frac{(-1)^{1/6} \sqrt{3}}{2^{1/3}} \, x + \frac{(-1)^{1/6} \sqrt{3}}{2^{1/3}} \, y  \, , \quad y \longmapsto -\left( -\frac{1}{2} \right)^{2/3} x + \left( -\frac{1}{2} \right)^{2/3} y
$$
such that $f(x^2y-y^3) = x^3-y^3$. In this way we find indecomposables $(\begin{smallmatrix}0&x\\ x^2 & 0\end{smallmatrix}) \otimes (\begin{smallmatrix}0&-y\\ y^2 & 0\end{smallmatrix})$ and 
$$
P_{j} = \begin{pmatrix}
0 & x-\eta^j y \\ \frac{x^d-y^d}{x-\eta^j y} & 0
\end{pmatrix}
, \quad
\eta = \E^{2\pi\I j/d}
\, , \quad
j\in \{ 0, \ldots, d-1 \}
$$
for $d=3$. These are all indecomposables of $\MF(x^3-y^3)$. We shall not discuss their deformations. 

\medskip

The case $d=4$ is more interesting: $\MF(x^4-y^4)$ is not of finite type, but it is \textsl{tame}, see e.\,g.~\cite{DrozdGreuelCMtype}. This means that indecomposable matrix factorisations of a given rank form a finite set of \textsl{1-parameter families} (together with a finite number of additional indecomposables). To our knowledge this is a (non-) existence result and the 1-parameter families have not been explicitly constructed. Below we will describe one such family that we found using the deformation method of section~\ref{subsec:naive}. 

Before we present the details of this family we would like to point out why it is of some interest from a CFT point of view. The Landau-Ginzburg model with potential $x^d-y^d$ (for any $d\geqslant 3$) is expected to have a CFT as its infrared fixed-point under RG flow whose symmetry algebra is the tensor product of two super Virasoro algebras. It is in fact a rational CFT, and hence there are only finitely many isomorphism classes of indecomposable boundary conditions that preserve the full symmetry algebra. In particular there cannot be any families of boundary conditions of this kind, and thus the family of matrix factorisations described below would correspond to boundary conditions that respect only part of the symmetry, namely that of a single superconformal algebra embedded in the tensor product. But for such branes the methods of rational CFT do not apply, and no such families have been constructed by other means. It would be interesting to understand our family from within the CFT perspective. 

As our starting point we take the matrix factorisation $(\begin{smallmatrix}0&y^2\\ -y^2 & 0\end{smallmatrix}) \otimes (\begin{smallmatrix}0&x\\ x^3 & 0\end{smallmatrix})$. Applying the method of section~\ref{subsec:naive} we obtain a deformed matrix factorisation that depends on four parameters $u_{i}$ before the obstruction constraints are imposed. One particular solution of these constraints sets all but one of the $u_{i}$ to zero while leaving the final parameter unconstrained. Denoting this parameter~$u$ and writing $\omega=\E^{\pi\I/4}$ we arrive at the following family of matrix factorisations (see~\cite{cdrMFdeform} for further details): 
\be\label{MFfamily}
Q(u) = 
\begin{pmatrix}
0 & 0 & y^2 & x - \omega u y \\
0 & 0 & -x^3 - \omega u x^2 y & \I u^2 x^2 - y^2 \\
\I u^2 x^2 - y^2 & -x + \omega u y & 0 & 0 \\
x^3 + \omega u x^2 y & y^2 & 0 & 0
\end{pmatrix} .
\ee

The fact that $Q(u)$ is not isomorphic to $Q(u')$ unless $u=u'$ can be seen as follows. For any matrix factorisation $Q=(\begin{smallmatrix}0&q_{0}\\ q_{1} & 0\end{smallmatrix})$ the associated \textsl{$t$-th Fitting ideal} is by definition generated by all $t$-minors of the matrix $q_{1}$. It is a general result, see e.\,g.~\cite[p.~21]{BrunsHerzogBook}, that 
the Fitting ideals of~$Q$ are invariants of~$Q$. This means that if two matrix factorisations have different Fitting ideals then they are not isomorphic. One easily checks that for different values of~$u$ the Fitting ideals of $Q(u)$ above do not agree, and we see that $Q(u)$ really is a 1-parameter family. 

\medskip

The category $\MF(x^d-y^d)$ is even more interesting for $d\geqslant 5$. In this case it is \textsl{wild}~\cite[Thm.~3]{DrozdGreuelCMtype}, which means that it is neither of finite type nor tame. In particular $\MF(x^d-y^d)$ then has many-parameter families, and it would be interesting to construct them explicitly. Applying the same deformation method as before leads to obstruction constraints with many solutions. Those few solutions that we studied in detail did however not give rise to families in the above strict sense.

\subsubsection*{Generalised permutation branes}

One may also study tensor products of distinct minimal models, i.\,e.~Landau-Ginzburg models with potential $W=x^d+y^{d'}$. If the exponents~$d$ and~$d'$ have a common factor there is a class of matrix factorisations of~$W$ known as generalised permutation branes~\cite{cfg0511078, fg0607095}. We will only consider the case $W=x^3+y^6$ where these branes are given by
$$
G_\omega = 
\begin{pmatrix}
0 & x - \omega y^2 \\
\frac{x^3+y^6}{x - \omega y^2} & 0
\end{pmatrix} ,
\quad
\omega^3 = -1 \, , 
$$
and their anti-branes. Every matrix factorisation of~$W$ that has been considered so far is a direct sum of generalised permutation branes and tensor product branes $(\begin{smallmatrix}0 & x^a \\ x^{3-a} & 0\end{smallmatrix}) \otimes (\begin{smallmatrix}0 & y^{2b} \\ y^{2(3-b)} & 0\end{smallmatrix})$. 

Since the fermionic self-spectra of generalised permutation branes are trivial (one easily computes $\dim\Ext^1(G_\omega,G_\omega)=4$ and $\dim\Ext^1(G_\omega,G_\omega)=0$), there are no directions into which~$G_\omega$ can be deformed. Instead we deform the superposition 
$$
Q = G_{\E^{\pi\I/3}} \oplus G_{-1}
$$
which has~8 bosons and~4 fermions. 

Both deformation algorithms from sections~\ref{subsec:naive} and~\ref{subsec:Massey} terminate at order~4 when applied to~$Q$. Most of the solutions of the associated quartic obstruction equations lead to matrix factorisations which are of type~$G_\omega$. Hence a direct sum of two generalised permutation branes can be deformed to a single such brane. However, one finds that there is one particular solution that gives rise to a deformation of~$Q$ which is not a direct sum of generalised permutation branes and tensor product branes. It has~6 bosons and~2 fermions and turns out to be isomorphic to the cone of the open string state
$$
\begin{pmatrix}
y & 0 \\
0 & xy - \frac{1}{1-\E^{-\pi\I/3}} y^3
\end{pmatrix} :
\bar Q_{\E^{\pi\I/3}} \longrightarrow Q_{-1} \, . 
$$

\subsubsection*{Simple branes on Calabi-Yau hypersurfaces}

Our final two examples of tensor product matrix factorisations are branes in Calabi-Yau hypersurfaces. These two examples have already been considered in~\cite{ks0812.2429} to which we refer for the explicit representatives of the boundary fields by which we deform below. 

First we look at 
$
Q = 
(\begin{smallmatrix}
0 & x_1^6 \\ x_1^6 & 0
\end{smallmatrix})
\otimes
(\begin{smallmatrix}
0 & x_2^6 \\ x_2^6 & 0
\end{smallmatrix})
\otimes
(\begin{smallmatrix}
0 & x_3^3 \\ x_3^3 & 0
\end{smallmatrix})
\otimes
(\begin{smallmatrix}
0 & x_4^3 \\ x_4^3 & 0
\end{smallmatrix})
\otimes
(\begin{smallmatrix}
0 & x_5 \\ x_5 & 0
\end{smallmatrix})
$. 
In~\cite[sect.\,4.3.1]{ks0812.2429} deformations (with non-zero obstructions) of this brane by the two bulk fields $x_1^6 x_2^6$ and $x_1 x_2 x_3 x_4 x_5$ as well as four boundary fields were computed up to fourth order. Using our implementation, we can go to order 50 in less than 40 minutes; in both cases the algorithm does not terminate at the given order. However, if only two particular boundary fermions are turned on, our algorithm terminates at order~4 without any obstructions, in agreement with~\cite[sect.\,4.3.3]{ks0812.2429}. 

The second example is 
$
Q = 
(\begin{smallmatrix}
0 & x_1^7 \\ x_1^7 & 0
\end{smallmatrix})
\otimes
(\begin{smallmatrix}
0 & x_2^3 \\ x_2^4 & 0
\end{smallmatrix})
\otimes
(\begin{smallmatrix}
0 & x_3^3 \\ x_3^4 & 0
\end{smallmatrix})
\otimes
(\begin{smallmatrix}
0 & x_4^3 \\ x_4^4 & 0
\end{smallmatrix})
\otimes
(\begin{smallmatrix}
0 & x_5 \\ x_5 & 0
\end{smallmatrix})
$. 
We consider deformations by the bulk fields $x_1^7 x_5$ and $x_1 x_2 x_3 x_4 x_5$, and again four boundary fields. Then our algorithm finds non-vanishing obstructions and terminates at order 7 after two minutes, consistent with~\cite[sect.\,6.3.1]{ks0812.2429}.

\subsubsection*{Other examples}

We have applied our code to an extensive list of further (mostly geometric) examples, including linear matrix factorisations~\cite{err0508053}, D0-branes on the quintic (see e.\,g.~\cite{j0612095}) and various D2-branes on Calabi-Yau hypersurfaces studied in~\cite{bbg0704.2666, bbpSOON}. We managed to compute deformations and obstructions to high orders, always consistent with known results and expectations, but to keep our presentation reasonably short we refrain from listing any details. From the point of view of studies like~\cite{bbg0704.2666, bbpSOON} only the binary information of whether deforming in a certain direction is obstructed or not is relevant; it should be interesting to include knowledge of the precise form of obstructions into such analyses.

\section{Some general observations}\label{sec:observations}

We conclude the paper with some general observations on how to generate new matrix 
factorisations and on how to deal with obstructions. While these remarks may seem 
obvious in hindsight, studying deformation algorithms and producing a large pool 
of concrete examples definitely helped recognise the patterns. 

\subsection{Nilpotent substitutions}\label{subsec:boosts}

\def\matu{\widehat{u}}
The $u_i$ in the deformed matrix factorisation $\qdef(x;u)$ are of course intended as complex 
parameters, but all the computations performed in any of the algorithms from section~\ref{sec:algorithms} remain correct as long as the $u_i$ commute among themselves and with the variables $x_l$ 
of the Landau-Ginzburg potential $W(x)$. This observation opens up an interesting playground: 
we can e.\,g.\ replace the $u_i$ by (commuting) matrices $\matu_i$ which solve the obstructions 
$f_j(\matu) = 0$. 

An especially simple choice is to use nilpotent matrices of size $N$, 
\be\label{nilpsubst1}
u_i \longmapsto \matu_i := v_i M_{[N]} \, , \quad v_i \in \C \, , \quad M_{[N]} \in {\rm Mat}_N(\C)
 \text{ such that } M_{[N]}^N = 0, 
\ee
for example $(M_{[N]})_{ab} = \delta_{a+1,b}$. 
This leads to a matrix factorisation $\qdef(x;\matu) \in {\rm Mat}_{2M}(\C[x]) 
\otimes {\rm Mat}_N(\C[v])$, where~$2M$ is the size of~$Q$; note that every $u_i^0$ 
is to be replaced by $\one_N$. 

The parameters $v_i$ are unconstrained as long as the nilpotency degree $N$ is chosen 
appropriately. (Since the obstructions have no linear term, $N=2$ is always possible.)
Thus we obtain a (higher rank) matrix factorisation of $W(x)$ without having to 
impose further restrictions to any zero locus. In particular, we will see that even 
if the zero locus of the deformation is trivial, $L_f = \{0\}$, usually 
$\qdef(x,\matu)$ is not equivalent to the original matrix factorisation $Q(x)$. 

\smallskip
Before looking at examples, let us note that nilpotent substitutions make connections 
between deformations as (iterated) cones visible. At nilpotency degree $N=2$, the 
substitution gives, up to re-ordering rows and columns, 
\be
\qdef(x,\matu) \cong
\begin{pmatrix}
Q(x) & \qd{1}(x;v)  \\
0 & Q(x) \\ 
\end{pmatrix} 
\ee
which is the cone $\text{C}(\psi)$ over the fermionic morphism $\psi=\qd{1}(x;v)$. 
At $N=3$, a ``double cone'' supported by $\text{C}(\psi)$ and $Q$ emerges, 
\be
\qdef(x,\matu) \cong
\begin{pmatrix}
Q(x) & \qd{1}(x;v) & \qd{2}(x;v) \\
0 & Q(x) & \qd{1}(x;v)  \\ 
0 & 0    & Q(x)   \\
\end{pmatrix} ,
\ee
and so on. This connection makes it plausible that, via nilpotent substitution, 
one can generate at least part of the category $\MF(W)$, just as one can via 
triangle-generation using (iterated) cones. 

\smallskip
For concreteness, let us look at one of the simplest possible examples, 
$W=x^5$ factorised as $x^5 = x\cdot x^4$. This matrix factorisation has 
one fermion and one boson, and deforming it leads to 
\be\label{simplestx5}
\qdef(x;u) = \begin{pmatrix}
0 & x+ u_{1} \\
x^4 -x^{3}u_{1}+x^{2}u_{1}^{2}-xu_{1}^{3}+u_{1}^{4} & 0
\end{pmatrix}
\ee
with $\qdef(x;u)^2 = (x^5 + u_1^5)\cdot \one_2$. When taken as 
as an equation for a complex parameter, $f(u)=u_1^5 = 0$ admits
the trivial solution $u_1=0$ only, but replacing $u_1$ by a rank 2 
nilpotent matrix $\matu$ makes the obstruction $f(\matu)$ disappear 
while $\qdef(x;\matu)$ is equivalent (by elementary row and column 
transformations) to the other elementary factorisation 
$x^5 = x^2\cdot x^3$ of the given Landau-Ginzburg potential, 
which has two fermions and two bosons. 

\smallskip
The minimal model associated to E$_6$ has a much richer category of matrix 
factorisations, with six indecomposable objects, but nevertheless we can 
generate a large part via nilpotent substitutions in a deformation of a 
single factorisation: with $W = W_{\textrm{E}_6}$ as above, take 
$$
Q = \begin{pmatrix} 0 & 0 & -x_2^2-x_3 & x_1  \\
          0 & 0 & x_1^2 & x_2^2-x_3  \\
      -x_2^2-x_3 & x_1 & 0 & 0  \\ 
   x_1^2 &-x_2^2+x_3   & 0 & 0  \\ 
\end{pmatrix} 
$$
which is $Q_5$ in the notation of~\cite{kst0511155}, 
and the simplest of the E$_6$ matrix factorisations. The Massey product 
formalism to deform was already applied in \cite{Siqveland2001} to 
deform $Q_5$ -- see also \cite{KnappOmer} --, and we refrain from repeating 
the explicit formula for $\qdef(x;u)$ here. Suffice it to say that there 
are two parameters $u_1$ and $u_2$, and that the zero locus of the obstructions
is trivial. 

On the other hand, using special nilpotent substitutions, we find (again using 
the numbering of branes from  \cite{kst0511155}) 
\begin{align}
\qdef(x;\matu) &\cong Q_6 \quad \textrm{for}\ \ \matu_1 = 0\,,\;\matu_2= M_{[2]}\,,
\nonumber\\ 
\qdef(x;\matu) &\cong Q_3 \quad \textrm{for}\ \ \matu_1 = M_{[2]}^T\,,\;\matu_2=0\,,
\nonumber\\ 
\qdef(x;\matu) &\cong Q_2 \quad \textrm{for}\ \ \matu_1 =  M_{[3]} + M_{[3]}^2 \,,\;\matu_2= 0\,,
\nonumber\\ 
\qdef(x;\matu) &\cong Q_4 \quad \textrm{for}\ \ \matu_1 = 0 \,,\;\matu_2=  M_{[4]}^2 + 
x_1\,M_{[4]}^3 + x_2\,E^{[4]}_{1,2} \,.
\nonumber
\end{align}
The nilpotent $N\times N$ matrices $M_{[N]}$ were defined after equation~\eqref{nilpsubst1}, 
and the $4 \times 4$ matrix $E^{[4]}_{1,2}$ in the last line has entries 
$(E^{[4]}_{1,2})_{ij}=\delta_{i,1}\delta_{j2}$. 
In these examples, checking for equivalences is best left to the computer ($Q_2$ is a 
$12\times12$ matrix), employing the method of~\cite{c0802.1624}. 
The only indecomposable E$_6$ matrix factorisation we have so far been unable to 
reproduce via nilpotent substitutions is the (relatively simple)~$Q_1$. 

\smallskip 

In the third and particularly the fourth case, we have already deviated from the 
simplest substitution pattern $u_i \longmapsto \matu_i = v_i\,M_{[N]}$ with the same 
nilpotent matrix $M_{[N]}$ for all $i$. 
It would be interesting to study the most general matrix substitutions one can 
use. In view of the algebra restrictions  (that the $\matu_i$ should commute and 
satisfy polynomial equations) this question should involve generalisations of 
Jordan-Chevalley decomposition and the Caley-Hamilton theorem. 

\smallskip

Let us add another small variation on the theme of re-interpreting the parameters $u_i$ 
as a way to generate matrix factorisations. Looking at \eqref{simplestx5}, one may 
recognise the transposition matrix factorisation $x^5+y^5= [x+y]\cdot[(x^5+y^5)/(x+y)]$ 
as first discussed in \cite{Ashok:2004zb}, but see also \cite{Brunner:2005fv}.
This is a matrix factorisation for a tensor product of two A-type minimal models, 
thus the bulk coordinate $y$ has a very different physical interpretation than the 
boundary parameter $u$ in \eqref{simplestx5} -- but at the level of formulas, they 
are but variables. One can exploit this ``bulk-boundary duality'' to produce 
deformed matrix factorisations for ``shorter'' Landau-Ginzburg polynomials from 
known matrix factorisations of ``longer'' ones:

Split the ``long'' polynomial as $W(x,y) = W_1(x) + W_2(x,y)$ with $W_1(x) = W(x,0)$, 
and assume that $\partial_{y_i}W_2(x,y)\big\vert_{y=0}=0$ for all $i$. If $Q(x,y)$ is a 
matrix factorisation of $W(x,y)$, we can set
$$
Q := Q(x,0)\quad \textrm{and}\quad Q_i :=  \partial_{y_i}Q(x,y)\big\vert_{y=0}
$$ 
and we automatically have $Q^2 = W_1\cdot\one$ and $\{ Q,Q_i\} = 0$. If $Q_i$ 
happens to be non-trivial in the fermionic cohomology of $Q$, it is a 
bona fide first order deformation with associated parameter $y_i$, and 
one obtains a deformed matrix factorisation of $W_1(x)$ for free: 
$\qdef(x;u) = Q(x,u)$. The term $W_2$ in the original potential $W$ 
plays the role of the total obstruction -- seemingly of a rather 
special form, but see the next section for further observations on this 
point. 

It should be interesting to explore connections between branes for 
different Landau-Ginzburg models further, in particular in view of the 
intriguing relation between bulk and boundary quantities.

\subsection{Lifting boundary obstructions by bulk deformations}\label{subsec:surjectivity}

There is one very surprising, and almost universal feature exhibited by the examples we have 
studied so far: the total obstructions $\obo$ to deforming a matrix factorisation~$Q$ of~$W$ 
are of the form 
\be\label{bulkobst}
\obo = f(x;u) \cdot \one \, , \quad f\in \Jac(W)[u] \, .
\ee
In words, the obstruction is proportional to the unit matrix, and the total obstruction 
polynomial $f(x;u)$ is a linear combination (with $u$-dependent coefficients) of elements 
in the Jacobi ring $\Jac(W)$ of the Landau-Ginzburg potential. 

In many of the cases where this is true, the chosen representatives for a basis of $\Ext^2(Q,Q)$ 
do have off-diagonal entries, thus some ``miraculous'' cancellation must be happening. 

\smallskip

The physical interpretation of this is as follows. On the one hand, since $\Jac(W)$ is the space of 
bulk states, whenever~\eqref{bulkobst} holds true, the obstructions against boundary deformations 
can be lifted by an appropriate perturbation $f(x;u)$ of the closed string background. On the other 
hand, since we have an explicit expression for $\qdef(x;u)$, we may read~\eqref{bulkobst} in the 
opposite direction: it shows how a brane~$Q$ in the background theory with potential~$W$ needs to 
change if it is to remain a brane under the perturbation of the background to $W+f(x;u)$. 

Whether the brane can ``survive'' arbitrary bulk perturbations, i.\,e.~whether the latter can be 
compensated by suitable boundary deformations, depends on whether all bulk fields occur in the 
obstruction and on the details of the $u$-dependent coefficients. 

\smallskip

Clearly it is desirable to have an easily verifiable criterion on 
$Q=(\begin{smallmatrix}0&q_{1}\\q_{0}&0\end{smallmatrix})$ which guarantees that the obstruction satisfies 
property~\eqref{bulkobst}, with the resulting interplay of bulk and boundary deformations. One can 
show that under certain assumptions, including that $\det(q_{1})=W$ and that~$W$ is irreducible, the 
bosonic cohomology $\Ext^2(Q,Q)$ is isomorphic to the Jacobi ring modulo some ideal, which 
in particular implies that every bosonic state in $\Ext^2(Q,Q)$ can be represented by a matrix of the 
form $f(x,u)\cdot \one$. 
However, in the majority of examples considered, the identity~\eqref{bulkobst} holds even if 
$\det(q_{1})$ does not divide~$W$, or if the very restrictive condition of~$W$ being irreducible 
is not met. 

A more general observation is that~\eqref{bulkobst} is always true if the bulk-boundary map
\be\label{beta}
\Jac(W) = \C[x_{1},\ldots,x_{n}]/(\partial W) \longrightarrow \Ext^2(Q,Q)
\ee
is surjective. The bulk-boundary map is part of the structure of open/closed topological field 
theory for Landau-Ginzburg models, and it simply maps~$\varphi$ to $\varphi\cdot \one$. Indeed, 
a general discussion of the possible surjectivity of bulk-boundary maps in topological field 
theories has already been initiated in~\cite{l0010269}, but to check whether~\eqref{beta} is 
surjective for a concrete Landau-Ginzburg boundary condition is relatively easy to do in practise: 
We choose bases $\varphi_{i}$ of $\Jac(W)$ and $\phi_{j}$ of $\Ext^2(Q,Q)$, then by reducing and 
lifting we can find a matrix~$C$ with polynomial entries such that
$$
\varphi_{i} \cdot \one = \sum_{j} C_{ij} \phi_{j} \, .
$$
If the rank of~$C$ equals the dimension of $\Ext^2(Q,Q)$ then~\eqref{beta} is surjective and the 
obstruction condition~\eqref{bulkobst} will hold. We have implemented this procedure as the function 
\texttt{Ext2check} in the library \texttt{MFdeform.lib} of~\cite{cdrMFdeform}. It is independent of 
any deformation algorithm, and can be easily applied to branes with geometric meaning, where 
our deformation procedures may not terminate. For example, the bulk-boundary map is surjective in 
the case of the D2 branes on the quintic studied e.\,g.~in~\cite{bbg0704.2666}, and it is not 
surjective for the D0 brane given in~\cite{j0612095}.

\subsection*{Acknowledgements}

We would like to thank C.~Baciu, Ch.~Barheine, I.~Burban, I.~Brunner, M.~Gaberdiel, B.~Hovinen, D.~Murfet, D.~Plencner, D.~Roggenkamp, E.~Scheidegger, M.~Soroush and in particular R.-O.~Buchweitz and M.~Kay for valuable discussions. This work was supported by STFC grant ST/G000395/1.


\begin{thebibliography}{10}

\bibitem{Ashok:2004zb}
  S.~K.~Ashok, E.~Dell'Aquila and D.~E.~Diaconescu,
  \textsl{Fractional branes in Landau-Ginzburg orbifolds},
  Adv.\ Theor.\ Math.\ Phys.\  {\bf 8} (2004), 461, 
 \href{http://www.arxiv.org/abs/hep-th/0401135}{[hep-th/0401135]}.

\bibitem{BrunsHerzogBook}
W.~Bruns and J.~Herzog, \textsl{Cohen-Macaulay rings}, Cambridge studies in advanced mathematics, Cambridge University Press, Cambridge U.K.~(1998). 

\bibitem{bbg0704.2666}
M.~Baumgartl, I.~Brunner, and M.~R.~Gaberdiel, \textsl{D-brane superpotentials and RG flows on the quintic}, JHEP \textbf{0707} (2007), 061,
  \href{http://arxiv.org/abs/0704.2666}{[arXiv:0704.2666]}.

\bibitem{bbpSOON}
M.~Baumgartl, I.~Brunner, and D.~Plencner, \textsl{D-brane Moduli Spaces and Superpotentials in a Two-Parameter Model}, to appear. 

\bibitem{Brunner:2005fv}
  I.~Brunner and M.~R.~Gaberdiel,
  \textsl{Matrix factorisations and permutation branes},
  JHEP {\bf 0507} (2005) 012
   \href{http://www.arxiv.org/abs/hep-th/0503207}{[hep-th/0503207]}.

\bibitem{Brunner:2005pq}
  I.~Brunner and M.~R.~Gaberdiel,
  \textsl{The matrix factorisations of the D-model},
  J.~Phys.~A  {\bf 38} (2005), 7901, 
   \href{http://www.arxiv.org/abs/hep-th/0506208}{[hep-th/0506208]}.

\bibitem{bhls0305}
I.~Brunner, M.~Herbst, W.~Lerche, and B.~Scheuner, \textsl{Landau-{G}inzburg
  {R}ealization of {O}pen {S}tring {TFT}}, JHEP \textbf{0611} (2003), 043,
  \href{http://www.arxiv.org/abs/hep-th/0305133}{[hep-th/0305133]}.

\bibitem{c0802.1624}
N.~Carqueville, \textsl{Triangle-generation in topological D-brane categories}, JHEP \textbf{04} (2008), 031,
  \href{http://arxiv.org/abs/0802.1624}{[arXiv:0802.1624]}.

\bibitem{c0904.0862}
N.~Carqueville, \textsl{Matrix factorisations and open topological string
  theory}, JHEP \textbf{0907} (2009), 005,
  \href{http://arxiv.org/abs/0904.0862}{[arXiv:0904.0862]}.
  
\bibitem{cdrMFdeform}
N.~Carqueville, L.~Dowdy, and A.~Recknagel, \emph{Code to compute deformations of matrix factorisations},
  \href{http://nils.carqueville.net/MFdeform}{http://nils.carqueville.net/MFdeform}.  

\bibitem{ck1104.5438}
N.~Carqueville and M.~M.~Kay, \textsl{Bulk deformations of open topological string theory}, 
  \href{http://arxiv.org/abs/1104.5438}{[arXiv:1104.5438]}.

\bibitem{cfg0511078}
C.~Caviezel, S.~Fredenhagen, and M.~R.~Gaberdiel, \textsl{The RR charges of A-type Gepner models}, JHEP \textbf{0601} (2006), 111,
  \href{http://arxiv.org/abs/hep-th/0511078}{[hep-th/0511078]}.

\bibitem{c0412149}
K.~J. Costello, \textsl{Topological conformal field theories and {C}alabi-{Y}au
  categories}, Adv.~ in Math.~\textbf{210} (2007),
  \href{http://arxiv.org/abs/math.QA/0412149}{[math.QA/0412149]}.

\bibitem{DrozdGreuelCMtype}
Y.~A.~Drozd and M.-G.~Greuel, \textsl{Cohen-Macaulay module type}, Compositio Mathematica \textbf{89} (1993), 315--338. 

\bibitem{err0508053}
H.~Enger, A.~Recknagel, and D.~Roggenkamp, \textsl{Permutation branes and linear
  matrix factorisations}, JHEP \textbf{0601} (2006), 087,
  \href{http://www.arxiv.org/abs/hep-th/0508053}{[hep-th/0508053]}.

\bibitem{fg0607095}
S.~Fredenhagen and M.~R.~Gaberdiel, \textsl{Generalised $N=2$ permutation branes}, JHEP \textbf{0601} (2006), 041,
  \href{http://arxiv.org/abs/hep-th/0607095}{[hep-th/0607095]}.

\bibitem{GreuelLossen}
G.-M.~Greuel, C.~Lossen, and E.~Shustin, 
   \textsl{Introduction to Singularities and Deformations}, Springer 2007.  

\bibitem{hll0402}
M.~Herbst, C.~I. Lazaroiu, and W.~Lerche, \textsl{Superpotentials, ${A}_{\infty}$
  {R}elations and {WDVV} {E}quations for {O}pen {T}opological {S}trings}, JHEP
  \textbf{0502} (2005), 071,
  \href{http://www.arxiv.org/abs/hep-th/0402110}{[hep-th/0402110]}.

\bibitem{hw0404196}
K.~Hori and J.~Walcher, \textsl{F-term equations near Gepner points}, JHEP \textbf{0501} (2005), 008,
  \href{http://www.arxiv.org/abs/hep-th/0404196}{[hep-th/0404196]}.

\bibitem{j0612095}
H.~Jockers, \textsl{D-brane monodromies from a matrix-factorization perspective}, 
JHEP \textbf{0702} (2007), 006,
  \href{http://arxiv.org/abs/hep-th/0612095}{[hep-th/0612095]}.

\bibitem{k0504437}
T.~Kadeishvili, \textsl{On the homology theory of fibre spaces}, Uspekhi Mat.
  Nauk \textbf{35:3} (1980), 183--188, English translation in Russian
  Math.~Surveys, 35:3 (1980), 231--238,
  \href{http://arxiv.org/abs/math.AT/0504437}{[math.AT/0504437]}.

\bibitem{kst0511155}
H.~Kajiura, K.~Saito, and A.~Takahashi, \textsl{Matrix Factorizations and Representations of Quivers II: type ADE case}, Adv. in Math. \textbf{211} (2007), 327--362, 
  \href{http://www.arxiv.org/abs/math.AG/0511155}{[math.AG/0511155]}.

\bibitem{kl0210}
A.~Kapustin and Y.~Li, \textsl{D-branes in {L}andau-{G}inzburg {M}odels and
  {A}lgebraic {G}eometry}, JHEP \textbf{0312} (2003), 005,
  \href{http://www.arxiv.org/abs/hep-th/0210296}{[hep-th/0210296]}.

\bibitem{KnappOmer}
 J.~Knapp and H.~Omer,
  \textsl{Matrix factorizations, minimal models and Massey products}, JHEP {\bf 0605} (2006), 064, 
   \href{http://arxiv.org/abs/hep-th/0604189}{[hep-th/hep-th/0604189]}.

\bibitem{ks0812.2429}
 J.~Knapp and E.~Scheidegger,
  \textsl{Matrix Factorizations, Massey Products and F-Terms for Two-Parameter Calabi-Yau Hypersurfaces}, Adv. Theor. Math. Phys. 14 (2010), 225--308, 
  \href{http://arxiv.org/abs/0812.2429}{[arXiv:0812.2429]}.
  
\bibitem{k9709040}
M.~Kontsevich, \textsl{Deformation quantization of {P}oisson manifolds}, Lett. Math.
  Phys. \textbf{66} (2003), 157--216,
  \href{http://arxiv.org/abs/q-alg/9709040}{[math.QA/9709040]}.

\bibitem{Laudal1983}
O.~A.~Laudal, \textsl{Matric Massey products and formal moduli I}, Lecture Notes in 
  Mathematics \textbf{1183} (1983), 218--240.

\bibitem{l0010269}
C.~I. Lazaroiu, \textsl{On the structure of open-closed topological field theory
  in two dimensions}, Nucl.~Phys.~B \textbf{603} (2001), 497--530,
  \href{http://www.arxiv.org/abs/hep-th/0010269}{[hep-th/0010269]}.

\bibitem{l0107162}
C.~I. Lazaroiu, \textsl{String field theory and brane superpotentials}, JHEP \textbf{0110} (2001), 018,
  \href{http://www.arxiv.org/abs/hep-th/0107162}{[hep-th/0107162]}.

\bibitem{l0312}
C.~I.~Lazaroiu, \textsl{On the boundary coupling of topological
  {L}andau-{G}inzburg models}, JHEP \textbf{0505} (2005), 037,
  \href{http://www.arxiv.org/abs/hep-th/0312286}{[hep-th/0312286]}.

\bibitem{lpwz0606144}
D.-M.~Lu, J.~H.~Palmieri, Q.-S.~Wu, and J.~J.~Zhang, \textsl{A-infinity structure on {E}xt-algebras},
  J.~Pure Appl.~Alg.~\textbf{213} (2009), no.~11, 2017--2037, 
  \href{http://arxiv.org/abs/math.KT/0606144}{[math.KT/0606144]}.

\bibitem{m9809}
S.~A. Merkulov, \textsl{Strongly homotopy algebras of a {K}\"{a}hler manifold},
  Internat. Math. Res. Notices \textbf{3} (1999), 153--164,
  \href{http://arxiv.org/abs/math.AG/9809172}{[math.AG/9809172]}.

\bibitem{m0001007}
S.~A. Merkulov, \textsl{Frobenius$_\infty$ invariants of homotopy {G}erstenhaber
  algebras, {I}}, Duke Math. J. \textbf{105} (2000), 411--461,
  \href{http://arxiv.org/abs/math/0001007}{[math.AG/0001007]}.

\bibitem{ms0609042}
G.~W. Moore and G.~Segal, \textsl{D-branes and {K}-theory in {2D} topological
  field theory}, \href{http://arxiv.org/abs/hep-th/0609042}{[hep-th/0609042]}.

\bibitem{Schless1968}
M.~Schlessinger, \textsl{Functors of Artin rings}, Transactions Amer.~Math.~Soc.\ \textbf{130} 
(1968), 208--222, 
\href{http://www.jstor.org/stable/1994967}{JSTOR/1994967}

\bibitem{Siqveland2001}
A.~Siqveland, \textsl{The method of computing formal moduli}, J.~of Algebra \textbf{241} 
(2001), 291--327 

\bibitem{v1991}
C.~Vafa, \textsl{Topological {L}andau-{G}inzburg {M}odels}, Mod.~Phys.~Lett.~A
  \textbf{6} (1991), 337--346.

\bibitem{Yoshinobook}
 Y.~Yoshino, \textsl{Cohen-Macaulay modules over Cohen-Macaulay rings}, London Mathematical 
Society Lecture Note Series, Cambridge University Press, Cambridge U.K.~(1990). 

\end{thebibliography}
\end{document}